\documentclass[aip,jap,reprint,nofootinbib,superscriptaddress]{revtex4-1}  

\usepackage{bm}
\usepackage{amsfonts} 
\usepackage{amssymb}
\usepackage{graphicx}   
\usepackage[caption=false]{subfig}
\usepackage{upgreek}

\begin{document}
\title{A simple model for calculating magnetic nanowire domain wall fringing fields}
\author{A. D. West}
\affiliation{Durham University, Department of Physics, South Road, Durham,
DH1 3LE, UK}
\author{T. J. Hayward}
\affiliation{University of Sheffield, Department of Materials Science and Engineering, Portobello Street,
Sheffield, S1 3JD, UK}
\author{K. J. Weatherill}
\affiliation{Durham University, Department of Physics, South Road, Durham,
DH1 3LE, UK}
\author{T. Schrefl}
\affiliation{Fachhochschule St. P\"{o}lten, Matthias Corvinus-Stra\ss e 15, 3100 St. P\"{o}lten, Austria}
\author{D. A. Allwood}
\affiliation{University of Sheffield, Department of Materials Science and Engineering, Portobello Street,
Sheffield, S1 3JD, UK}
\author{I. G. Hughes}
\affiliation{Durham University, Department of Physics, South Road, Durham,
DH1 3LE, UK}

\begin{abstract}
We present a new approach to calculating magnetic fringing fields from head-to-head type domain walls in planar magnetic nanowires. In contrast to calculations based on micromagnetically simulated structures the descriptions of the fields are for the most part analytic and thus significantly less time and resource intensive. We begin with an intuitive picture of domain walls, which is built upon in a phenomenological manner. The resulting models require no \emph{a priori} knowledge of the magnetization structure, and facilitate calculation of fringing fields without any free parameters. Comparisons with fields calculated using micromagnetic methods show good quantitative agreement.
\end{abstract}
\maketitle

\section{Introduction}
\begin{figure}[h]
\center
\scalebox{0.95}{\includegraphics{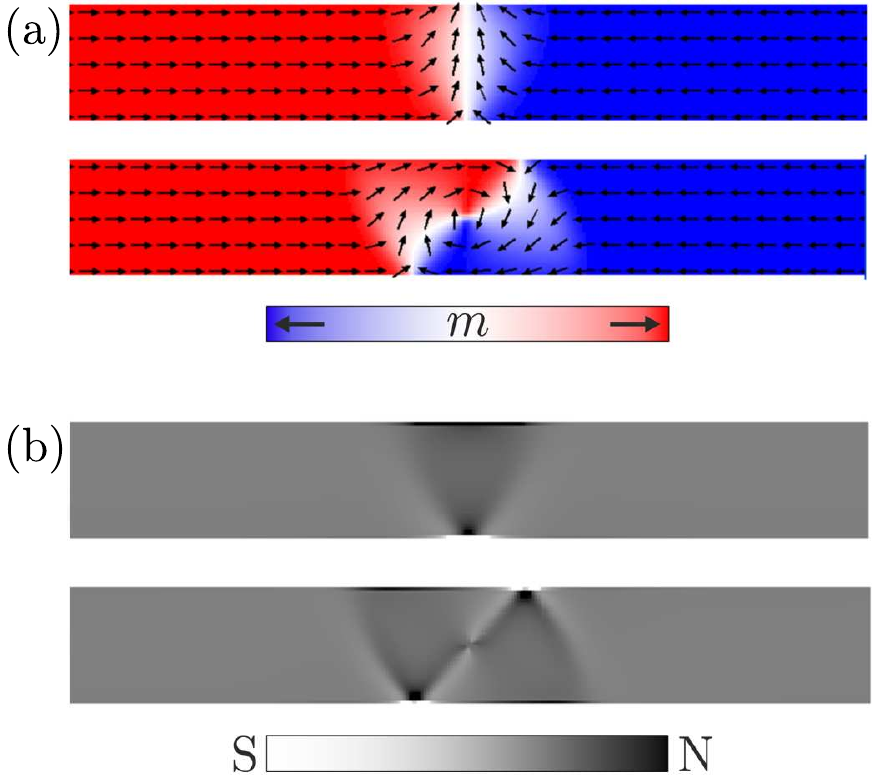}}
\caption{\label{fig:dw}(a) Magnetization structures of typical head-to-head domain walls as calculated via micromagnetic methods using OOMMF \cite{OOMMF}. The top image shows a transverse-type wall (200~nm $\times$ 5~nm wire cross section), the bottom image is a vortex-type wall (200~nm $\times$ 15~nm wire cross section). Shading represents the longitudinal component of the normalized magnetization, $m$. Arrows represent the in-plane magnetization direction. (b) Divergence of the magnetization. Dark (light) areas signify positive (negative) $\vec{\nabla}.\vec{M}$, or effective `north' (`south') poles. Cross sections are as in (a).}
\end{figure}

Magnetic domain walls are boundaries between areas of differing magnetization direction, with magnetization rotating through the domain wall width. In patterned magnetic nanowires, domain walls separate opposite magnetizations, either in a head-to-head (magnetizations point towards the wall) or tail-to-tail (magnetizations point away from the wall) configuration. The domain walls form in two characteristic shapes -- transverse or vortex \cite{dwshape1,dwshape2}. The former is more common in wires with smaller cross sections. The magnetization structures of typical domain walls are shown in Figure~\ref{fig:dw}, as calculated via micromagnetic methods. The converging or diverging magnetization of a domain wall results in an associated effective magnetic monopole moment. Thus, strong magnetic fields are found in close proximity to the domain walls. Directly above the walls these fields are directed out of the nanowire's plane.

The characteristics and dynamics of nanomagnetic domain walls have been the subjects of much research, e.g.\ \cite{naturematerials} and are key topics in the burgeoning field of spintronics \cite{spintronics,parkin}. The fields of atom optics and atom chips have also experienced rapid expansion and continue to strive towards improved techniques for producing arenas for atomic physics experiments and quantum information processing \cite{hindshughes,fortzimm,adamsoptics}. Within these research fields nanomagnetic domain walls have a number of possible applications, such as an atom mirror \cite{nanomirror} or mobile atom traps \cite{nanotrap}. A factor common to all these areas of study is the need for an accurate knowledge of the magnetic fringing fields generated by domain walls. In particular, for atomic physics applications a knowledge of the fields for distances of around 100~nm or greater is desired; at distances closer than this the attractive van der Waals interaction begins to modify significantly the atomic potential \cite{mohapatra}.

Rigorous micromagnetic methods currently allow accurate computation of magnetization configuration, and hence fields, at all distances. However they are resource and time intensive and require detailed understanding of micromagnetics. We present a more direct method of field calculation -- the `monopole model', which affords considerable perspicacity and utility, particularly in the regimes of interest for applications of domain walls in atomic physics experiments. In Section~\ref{sec:mono} we outline the models used, extending to greater complexity in a phenomenological manner. We discuss how emulating the detailed magnetization structure of the wall is not a good way of quickly and accurately computing these fields; the models we present do not demand this and can be used without free parameters. Analysis of the models' accuracy is described in Section \ref{sec:comparisons}.

\section{Models}
\label{sec:mono}
\begin{figure*}[!ht]
\centering
\subfloat[][A domain wall represented as a region of magnetization reversal. In the limit $s\to 0$ the wall is a 2D object of area $wt$.]{\includegraphics[scale=0.34]{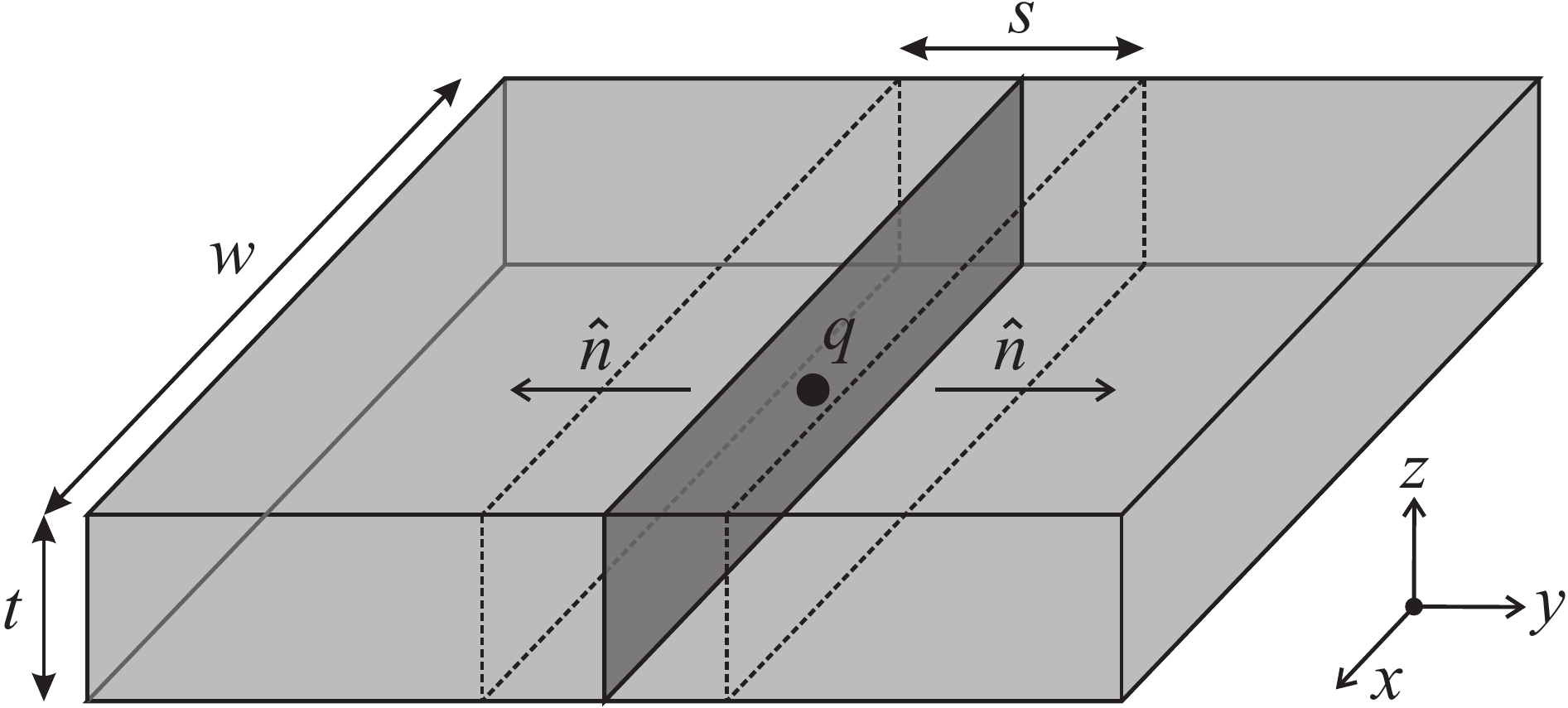}\label{fig:dwcharge}}
\qquad
\subfloat[][A 1D line of magnetic charge. Here the wire length is into the page.]{\includegraphics[scale=0.34]{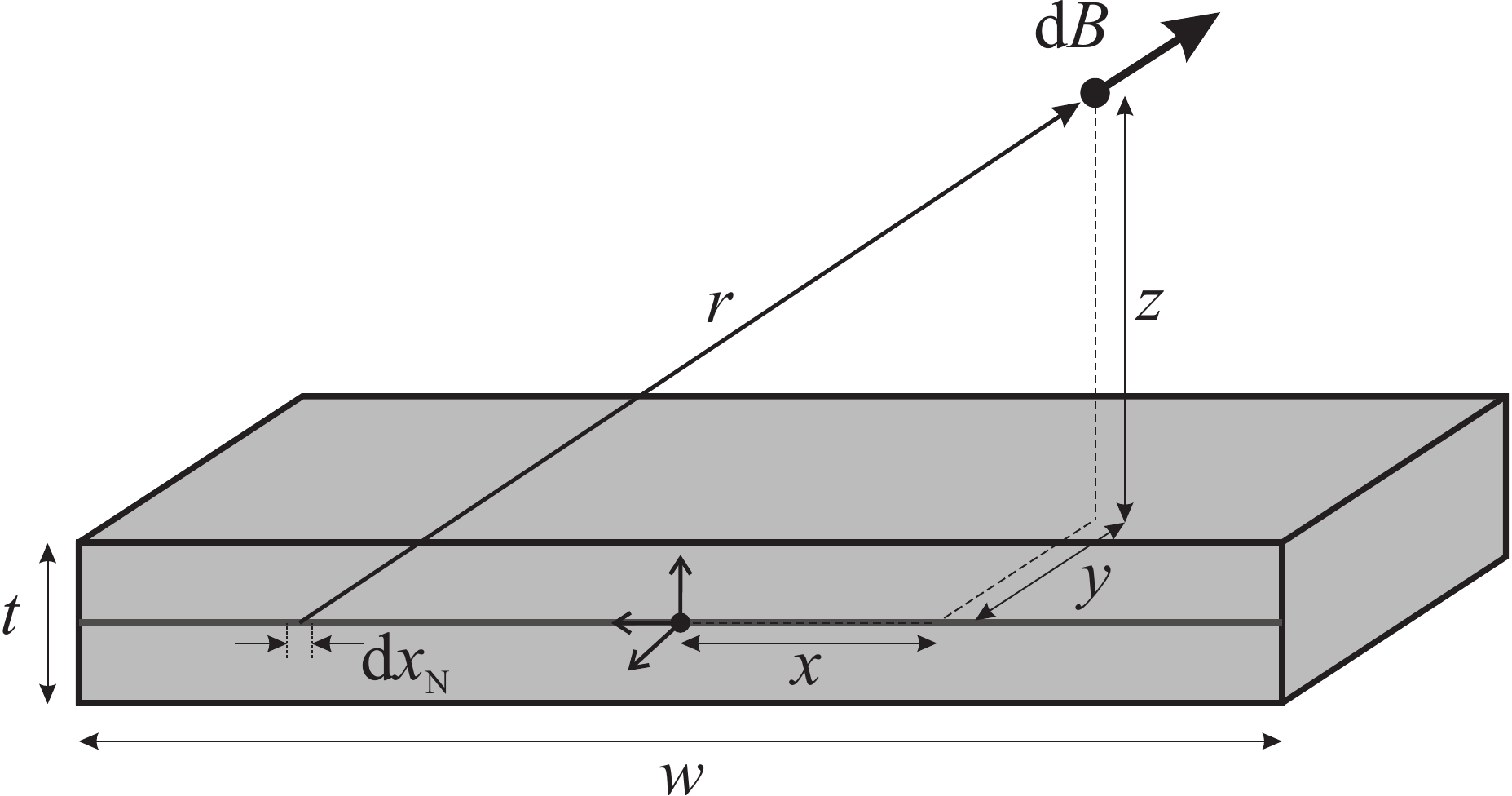}\label{fig:1d}}\\
\subfloat[][A 2D area of magnetic charge with finite size along the wire length.]{\includegraphics[scale=0.34]{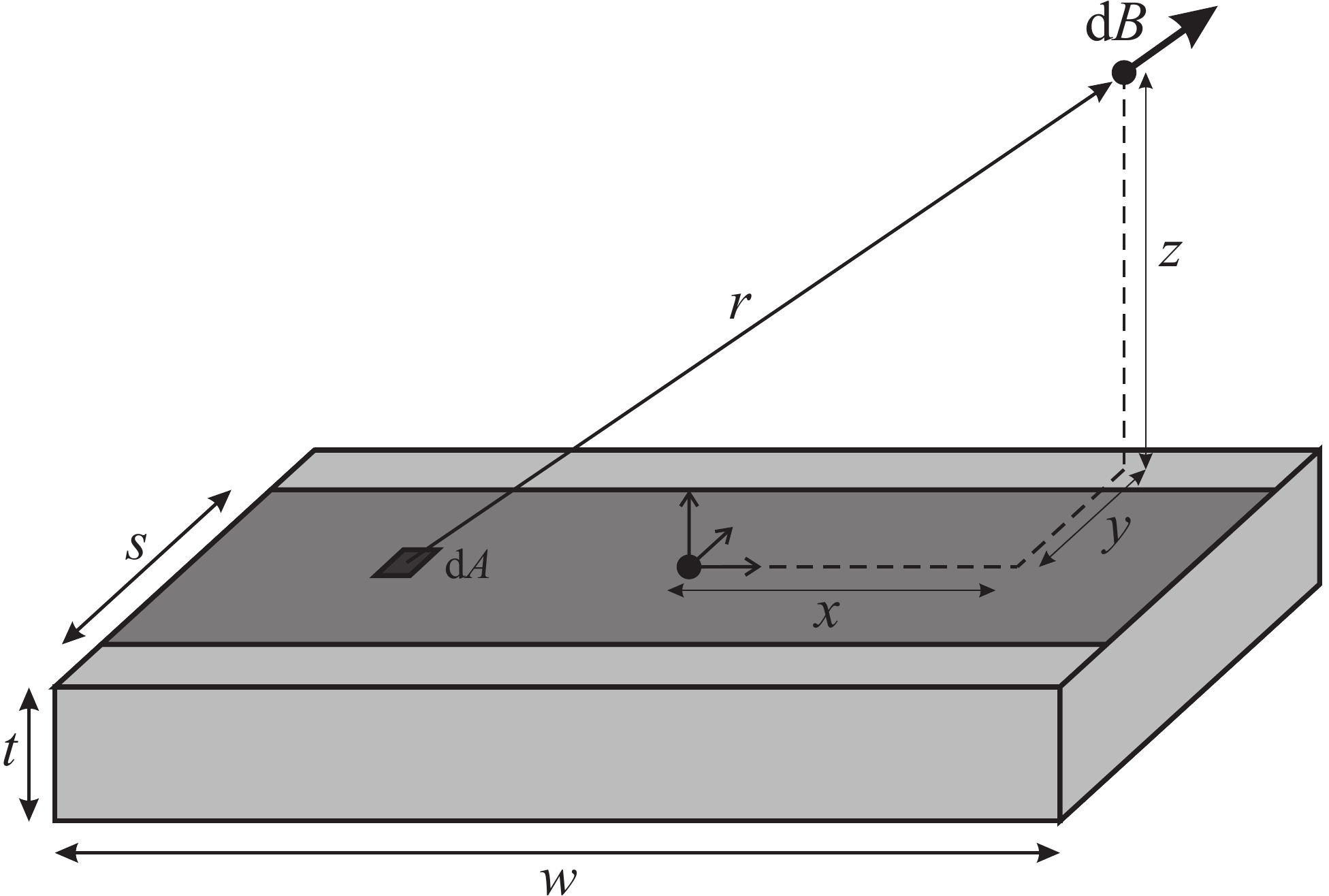}\label{fig:2db}}
\qquad
\subfloat[][A triangle of magnetic charge. This shape is applicable for transverse walls.]{\includegraphics[scale=0.34]{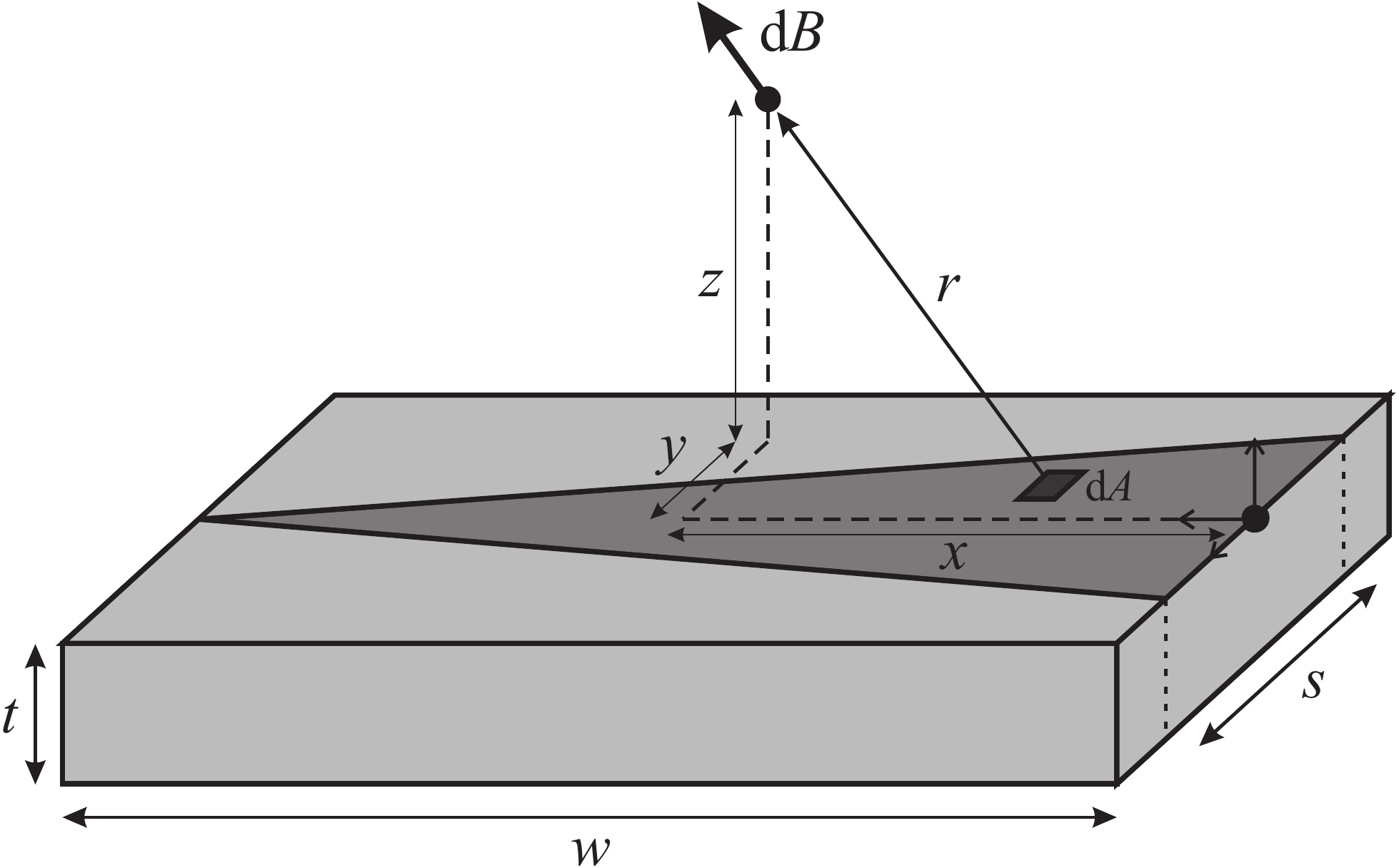}\label{fig:tri}}
\caption{Illustrations of the geometries of the four different models of the magnetic charge associated with a domain wall.}
\label{fig:4mods}
\end{figure*}

Illustrations of the four different models considered are shown in Figure \ref{fig:4mods}. The basic `monopole model' is described in Section \ref{subsec:monomodel} and more detailed models are developed in subsequent sections. Derivations are provided in Appendix~\ref{sec:fieldexp}.

\subsection{Monopole Model}
\label{subsec:monomodel}
Although Maxwell's equations preclude the existence of magnetic monopoles, they have been posited and observed as quasiparticles \cite{castelnovo,giblin}. We will use them as a theoretical construct; at large distances, nanomagnetic domain walls can be well approximated as point magnetic charges. This point acts as a source or sink of magnetic field lines -- a monopole. This prescription of an effective charge is a treatment followed elsewhere, see e.g. \cite{castelnovo,ladak,pinning}. The charge is then given by
\begin{equation}
q_{\rm m}=2\mu_0M_{\rm s}wt,
\end{equation}
where $M_{\rm s}$ is the saturation magnetisation, $\mu_0$ is the permeability, $w$ is the wire width and $t$ is the wire thickness. This expression has previously been derived\cite{nanomirror} -- the details are provided in Appendix~\ref{sec:fieldexp}. The magnetic flux density (`magnetic field' from hereon) for a charge $q_{\rm m}$ has an inverse square dependence on distance:
\begin{equation}
\vec{B}(\vec{r})=\frac{q_{\rm m}}{4\pi\left|\vec{r}\right|^2}\hat{r},
\end{equation}
where $\hat{r}$ is the unit vector associated with $\vec{r}$. Note that $\vec{B}$ and $\vec{H}$ differ only by a factor of $\mu_{\rm 0}$.

As will be shown in Section \ref{sec:comparisons}, this model is an excellent approximation at distances that are large compared to the width of the nanowire. We now extend the model to emulate more faithfully the shape of a real domain wall. This will significantly extend the region over which the model accurately reproduces the shape of real magnetic fields.

\subsection{1D Domain Wall}
\label{subsec:1d}
The first extension to the model is to consider the wall as a 1D object, i.e.\ a line of charge across the width of the wire as illustrated in Figure~\ref{fig:1d}. The corresponding expressions for the field are given in Eqs.~(\ref{eq:1dbx})-(\ref{eq:1dbz}). These expressions will be shown to offer a significant improvement in the accuracy of field calculations compared to the simple monopole model.

\subsection{2D Domain Wall}
\label{subsec:2d}
We now extend our model to consider the domain wall to be 2D as a more exact representation of it as an extended object. Modelling a domain wall with finite size in $z$ does not give significant improvement; the nanowire thickness is small compared to its width. From hereon we consider domain walls with zero size in $z$, but finite extent in $x$ and $y$, as illustrated in Figure~\ref{fig:2db}. The corresponding expressions for the magnetic field are shown in Eqs.~(\ref{eq:rectbx})-(\ref{eq:rectbz}).

Note that Equations~(\ref{eq:rectbx}) and (\ref{eq:rectby}) are identical under an exchange of $\left\{x,w\right\}$ with $\left\{y,s\right\}$, as expected due to symmetry.

\begin{widetext}
\begin{eqnarray}
B_x&=\frac{\mu_{\rm 0}M_{\rm s}t}{2\pi}\left[\frac{2}{\sqrt{(w-2x)^2+4\left(y^2+z^2\right)}}- \frac{2}{\sqrt{(w+2x)^2+4\left(y^2+z^2\right)}}\right]\label{eq:1dbx}\\
B_y&=\frac{\mu_{\rm 0}M_{\rm s}ty}{\pi(y^2+z^2)}\left[\frac{w-2x}{2\sqrt{(w-2x)^2+4\left(y^2+z^2\right)}}+ \frac{w+2x}{2\sqrt{(w+2x)^2+4\left(y^2+z^2\right)}}\right]\label{eq:1dby}\\
B_z&=\frac{\mu_{\rm 0}M_{\rm s}tz}{\pi(y^2+z^2)}\left[\frac{w-2x}{2\sqrt{(w-2x)^2+4\left(y^2+z^2\right)}}+ \frac{w+2x}{2\sqrt{(w+2x)^2+4\left(y^2+z^2\right)}}\right]\label{eq:1dbz}
\end{eqnarray} 

\begin{eqnarray}
B_x=&\frac{\mu_{\rm 0}M_{\rm s}t}{2\pi s}\log\left[\frac{(s-2y)+\sqrt{(2x-w)^2+(2y-s)^2+4z^2}} {(s-2y)+\sqrt{(2x+w)^2+(2y-s)^2+4z^2}}.\frac{(-s-2y)+\sqrt{(2x+w)^2+(2y+s)^2+4z^2}} {(-s-2y)+\sqrt{(2x-w)^2+(2y+s)^2+4z^2}}\right]\label{eq:rectbx}\\
\textstyle B_y=&\frac{\mu_{\rm 0}M_{\rm s}t}{2\pi s}\log\left[\frac{(w-2x)+\sqrt{(2x-w)^2+(2y-s)^2+4z^2}} {(w-2x)+\sqrt{(2x-w)^2+(2y+s)^2+4z^2}}.\frac{(-w-2x)+\sqrt{(2x+w)^2+(2y+s)^2+4z^2}} {(-w-2x)+\sqrt{(2x+w)^2+(2y-s)^2+4z^2}}\right]\label{eq:rectby}\\
\textstyle B_z=&\frac{\mu_{\rm 0}M_{\rm s}t}{2\pi s}\left\{\tan^{-1}\left[\frac{2(w/2-x)(s/2-y)}{z\sqrt{(2x-w)^2+(2y-s)^2+4z^2}}\right]+ \tan^{-1}\left[\frac{2(w/2-x)(s/2+y)}{z\sqrt{(2x-w)^2+(2y+s)^2+4z^2}}\right]\right.\nonumber\\
\textstyle\hspace{27pt}&-\left.\tan^{-1}\left[\frac{2(-w/2-x)(s/2-y)}{z\sqrt{(2x+w)^2+(2y-s)^2+4z^2}}\right]- \tan^{-1}\left[\frac{2(-w/2-x)(s/2+y)}{z\sqrt{(2x+w)^2+(2y+s)^2+4z^2}}\right]\right\}\label{eq:rectbz}
\end{eqnarray}
\end{widetext}

\subsection{Triangular Domain Walls}
Considering the magnetic pole distributions, $\vec{\nabla}.\vec{M}$, shown in Figure~(\ref{fig:dw}), there is a clear difference in the shape associated with the two wall types; whilst the poles of a vortex wall are contained within an approximately rectangular section of the nanowire, the pole distribution of a transverse wall is distinctly triangular. The final extension of our model is to incorporate this characteristic shape, so as to better emulate the field from a transverse wall. This situation is illustrated in Figure~\ref{fig:tri}.

Equations~(\ref{eq:dq1d}) and (\ref{eq:dintrect}) can be modified to incorporate a triangular shape. Unfortunately it was not possible to derive closed expressions for $B_z$ or $|\vec{B}|$ as with previous models. Whilst numerical integration is quicker than using micromagnetically simulated structures for the same level of resolution, it is significantly slower than previous analytic expressions. Instead, the computation can be sped up considerably by utilizing the result for a rectangular wall. One can approximate the shape of the triangular wall by a series of rectangles and sum the resulting field from each. Computing the field over $10^6$ points takes around 2 seconds per rectangle. Negligible loss of accuracy was observed using a triangle divided into 40 rectangles.

Whilst the models presented thus far emulate the basic shape of the domain walls' volume charge distributions, they do not take into account the positive and negative regions of edge charge present in both transverse and vortex walls (Figure~\ref{fig:dw}b). As can be seen in Figure~\ref{fig:shortlongrangefield}, at extremely short range the fields produced by the domain walls mimic this complex charge structure. A number of efforts were made to emulate these more subtle features; trapezoidal walls, dominating edge charge regions and simple spatial variations in charge density/polarity were all investigated. However, none of these approaches produced overall improvements in accuracy and therefore we do not describe them in detail here. Thus we conclude, in agreement with previous findings\cite{dwimaging}, that the overall shape of a domain wall's volume charge is by far the most important feature at all but the very shortest distances. Whilst the accuracy of the models in the near field could undoubtedly be improved by incorporating a detailed magnetization structure, it is precisely this we are trying to circumvent with the development of these simple models.

In the 2D models we use the approximation $s=w$. It is possible to optimize the value of $s$ through a comparison with micromagnetic simulations, however this approach is contrary to the aim of these models and significantly detracts from their utility. We will see later that using an optimized value of $s$ confers only a very small improvement in accuracy. Unless explicitly stated otherwise, $s=w$ will be used throughout. We note for a triangular wall (as with a rectangular one) that the distribution of charge across $x_{\rm N}$ is independent of $s$:
\begin{equation}
{\rm d}q_{\rm m}(x_{\rm N})=4M_{\rm s}\mu_{\rm 0}t/s\ {\rm d}A=4M_{\rm s}\mu_{\rm 0}tx_{\rm N}/w\ {\rm d}x_{\rm N}.\nonumber
\end{equation}
This is equivalent to the fact that all triangles have centers of mass at barycentric coordinates (1/3, 1/3, 1/3). The field maximum according to our analytic model is thus found at $\left(x,y\right)=\left(w/3,0\right)$ when sufficiently far from the wall. Note also integrating ${\rm d}q_{\rm m}(x_{\rm N})$ with respect to $x_{\rm N}$ yields Equation~(\ref{eq:charge}).

\section{Comparison of models}
\label{sec:comparisons}
\begin{figure*}[!t]
\centering
\scalebox{0.55}{\includegraphics{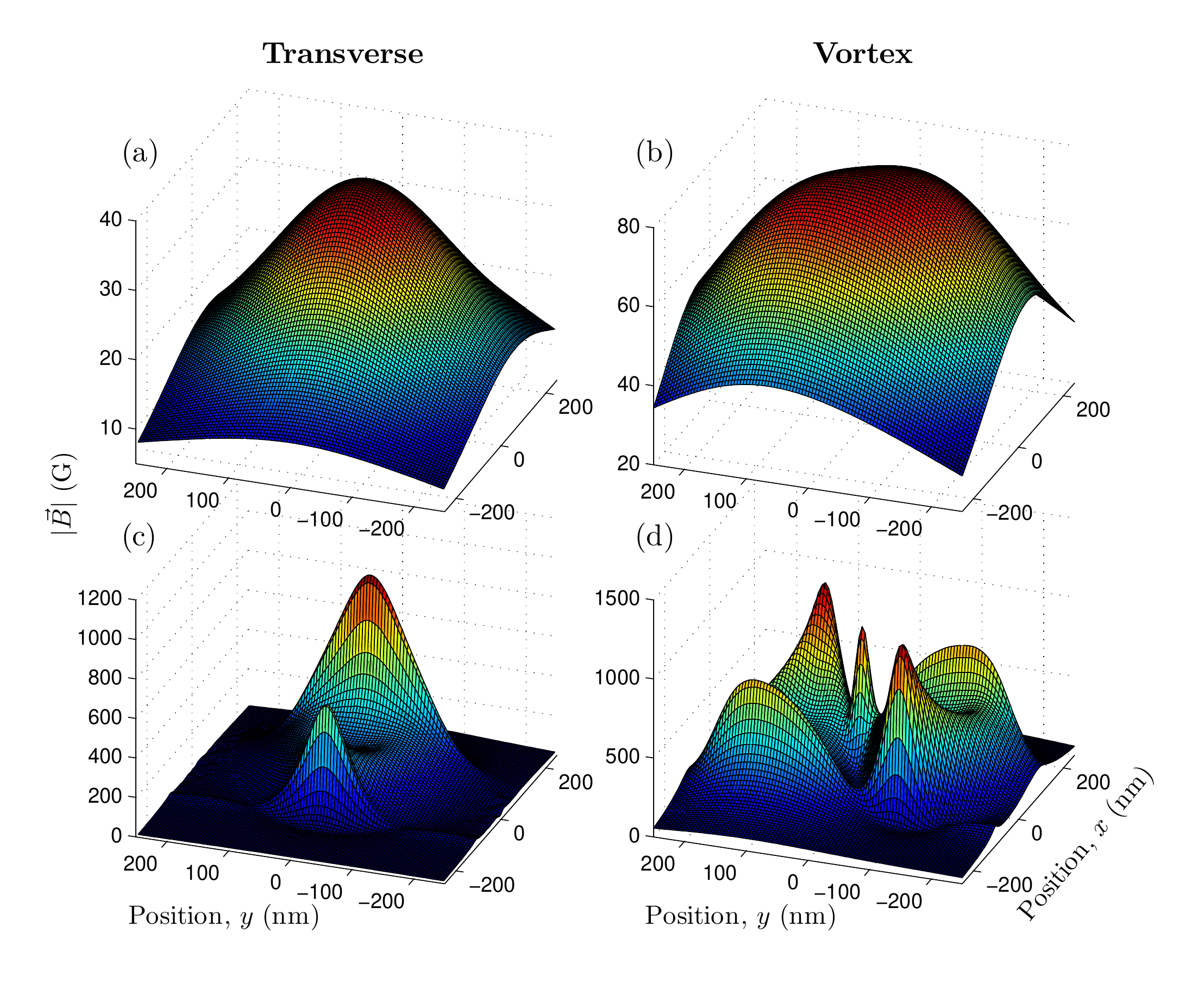}}
\caption{\label{fig:shortlongrangefield}Fields calculated via micromagnetic simulations. (a) and (b) are at $z=200~{\rm nm}$, whilst (c) and (d) are at $z=12.5~{\rm nm}$; $z=0$ is located in the center of the wire's thickness. (a) and (c) correspond to a transverse-type domain wall (cross section = 200~nm $\times$ 5~nm), whilst (b) and (d) correspond to a vortex-type wall (cross section = 200~nm $\times$ 15~nm). At short distances the field adopts a complex shape, indicative of the magnetic charge distribution. Further away the field is significantly smoother and smaller in magnitude. Note in the far field the shape is skewed in $x$ for the transverse domain wall.}
\end{figure*}

To assess the accuracy of the models presented comparison was made with results from calculations based on micromagnetic simulations. We use a proprietary micromagnetic code that solves the Landau-Lifshitz-Gilbert equation of motion and quasistatic Maxwell equations within a finite element/boundary element framework \cite{schrefl}. We simulate 8.4~$\upmu$m long nanowire structures, discretized within the domain wall region into tetrahedral meshes with a 5~nm characteristic size. Physically appropriate domain wall structures are introduced by imposing simple bi-domain states and then allowing relaxation to equilibrium. The magnetic field profiles above the domain walls are calculated analytically from the equivalent dipole charges on the nanowires' surfaces \cite{rech,nanotrap}. Magnetic fields created by effective magnetic charges at the nanowire ends are subtracted from the data using the simple point monopole approximation, which, as we will show later, is extremely accurate in the far-field. Figure~\ref{fig:shortlongrangefield} shows the increase in compexity of the field shape at short distances. Because of this complex shape the models presented here will always experience inaccuracy when calculating fields at very short distances.

We consider the fields calculated by all four models, for a transverse-type domain wall in a wire with $w=200$~nm and $t=5$~nm. A saturation magnetization, $M_{\rm s}$, of $8.6\times 10^5$~A/m was used throughout\cite{ferromagnetism}. The models were examined over a $1~\upmu$m$\ \times\ 1~\upmu$m$\ \times\ 1~\upmu$m cube divided evenly into $10^6$ points, with the domain wall centered at the bottom of the cube. This is representative of distances within which atomic physics applications of domain walls aim to work; outside this region the field is less than $1$~G.

\begin{figure}[!h]
\center
\hspace{-4pt}\scalebox{0.38}{\includegraphics{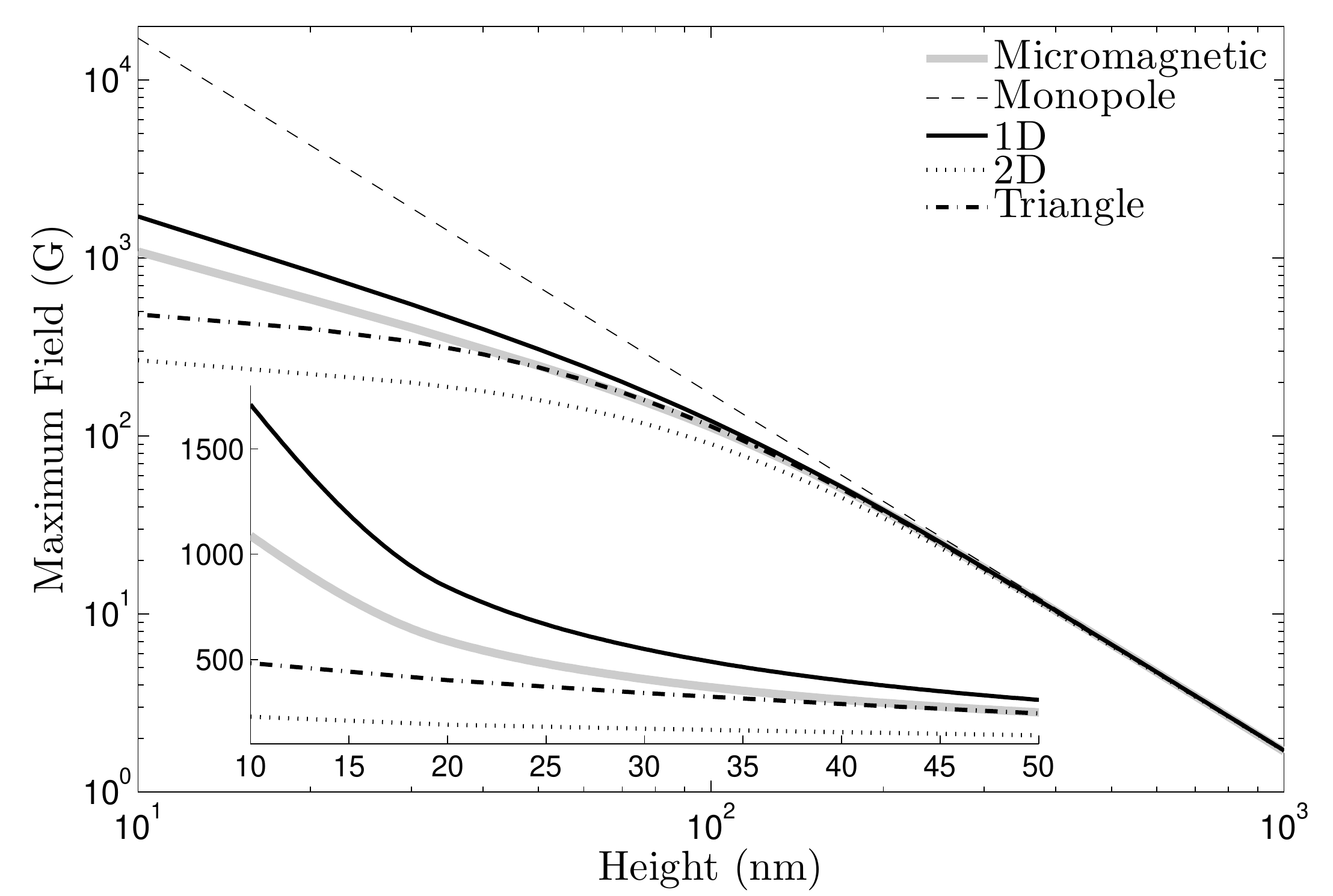}}
\caption{\label{fig:maxfieldcomparison}Maximum field at a given height above a transverse domain wall (200~nm $\times$ 5~nm wire) as calculated by the various models. Note the good agreement at $\gtrsim$50~nm for all models except the simple monopole model.}
\end{figure}

The following analysis examines only the field magnitude for the sake of brevity. Very good accuracy was also observed for the field direction and is discussed briefly towards the end of the section. A summary of the different figures of merit is provided in Appendix \ref{sec:fom}. Initial comparison was made by considering the maximum field at a given height, shown in Figure~\ref{fig:maxfieldcomparison}. This is an important quantity in atomic physics applications of domain walls, and shows well how the models scale with distance. 

From Figure~\ref{fig:maxfieldcomparison} it can be seen that at heights greater than 50~nm there is very good agreement ($<10\%$ error) with micromagnetic simulations from all but the simplest `monopole' model. There is a stark improvement moving to the 1D case - for $z\lesssim 300$~nm there is an order of magnitude improvement in the error. Above this height all the models converge with the micromagnetic simulations, having an error of less than 5\%. Extension beyond the 1D model does not afford significantly more accuracy.

\begin{figure}[h]
\center
\hspace{-5pt}\scalebox{0.38}{\includegraphics{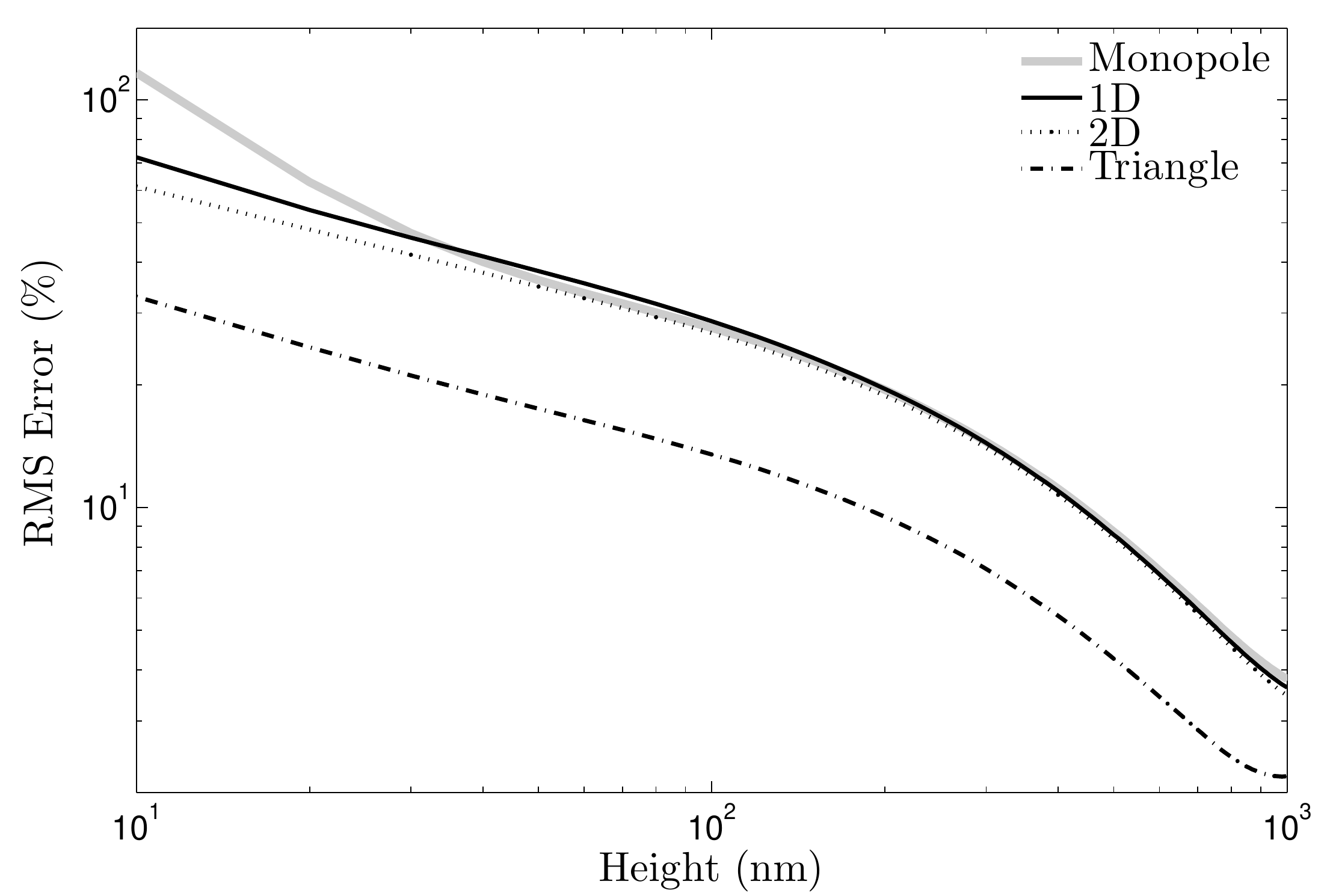}}
\caption{\label{fig:rmserrorcomparison}RMS accuracy of the field over all points at a given height above a transverse domain wall (200~nm $\times$ 5~nm wire). Note the monopole model is very inaccurate at shorter distances, and that incoporating a triangular shape confers significantly increased accuracy.}
\end{figure}

A more thorough analysis of the models examines the field over the entire region of space, so the shape of the field is tested. Figure~\ref{fig:rmserrorcomparison} shows the RMS error over all points at a given height for the different models considered. 
We see that there is a big improvement in adopting a triangular shape; more accurately imitating the shape of the domain wall produces a more accurate field shape. Typical values of the RMS and mean percentage errors are also provided in Appendix~\ref{sec:fom}. 

As can be seen in Figure~\ref{fig:rmserrorcomparison}, the accuracy of all the models is much worse at small distances, due to the dramatic change in the shape of the fields. An examination of e.g. Figure~\ref{fig:shortlongrangefield} suggests highly localised regions of magnetic charge for both wall types. However, as previously mentioned, improved accuracy is not gained by using a simple representation of these charge distributions. Whilst very good accuracy is achieved for all wire geometries at heights \textgreater 100~nm, some spintronics applications require knowledge of the field very close to the wire, e.g. read/write heads can be located at flying heights of $\sim$10~nm \cite{schrefl}. This is within the regime where these analytic models break down. 

These initial figures of merit are strong indicators of the utility of the model we present, allowing non-expert users to calculate fields at all but the closest distances with a very good level of accuracy. Whilst mean errors indicate well the overall accuracy of models, there may be small regions of space where very large errors exist. As such, maximum errors are also provided: $E_{\rm Max}$ is the maximum \% error in the field, $E_{\rm Max}^{>100}$ is the maximum \% error above 100~nm. $\Delta B_{\rm Max}$ and $\Delta B_{\rm Max}^{>100}$ are the corresponding absolute errors in the field. A summary of the figures of merit is given in Appendix~\ref{sec:fom}.

\begin{figure}
\centering
\includegraphics[scale=1.7]{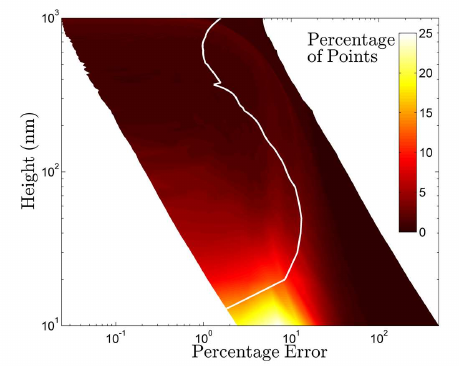}
\caption{The distribution of the percentage error in the field at all heights as calculated by the triangular model for a $200~{\rm nm} \times 5~{\rm nm}$ nanowire. At each height the data are grouped into 100 evenly distributed bins. The white line shows the error directly above the wall barycenter.}
\label{fig:errordist}
\end{figure}

The best overall representation of the accuracy of the models is achieved by examining the distribution of the error. This is shown for the 200~nm $\times$ 5~nm wire, using the triangle model, in Figure \ref{fig:errordist}. At all heights the large majority of points have an error $<20\%$. The error directly above the triangle barycenter (white line) is quite high within the points at a given height; this is the point closest to the wall for a given height.

\begin{figure}[h]
\center
\scalebox{0.39}{\hspace{-4pt}\includegraphics{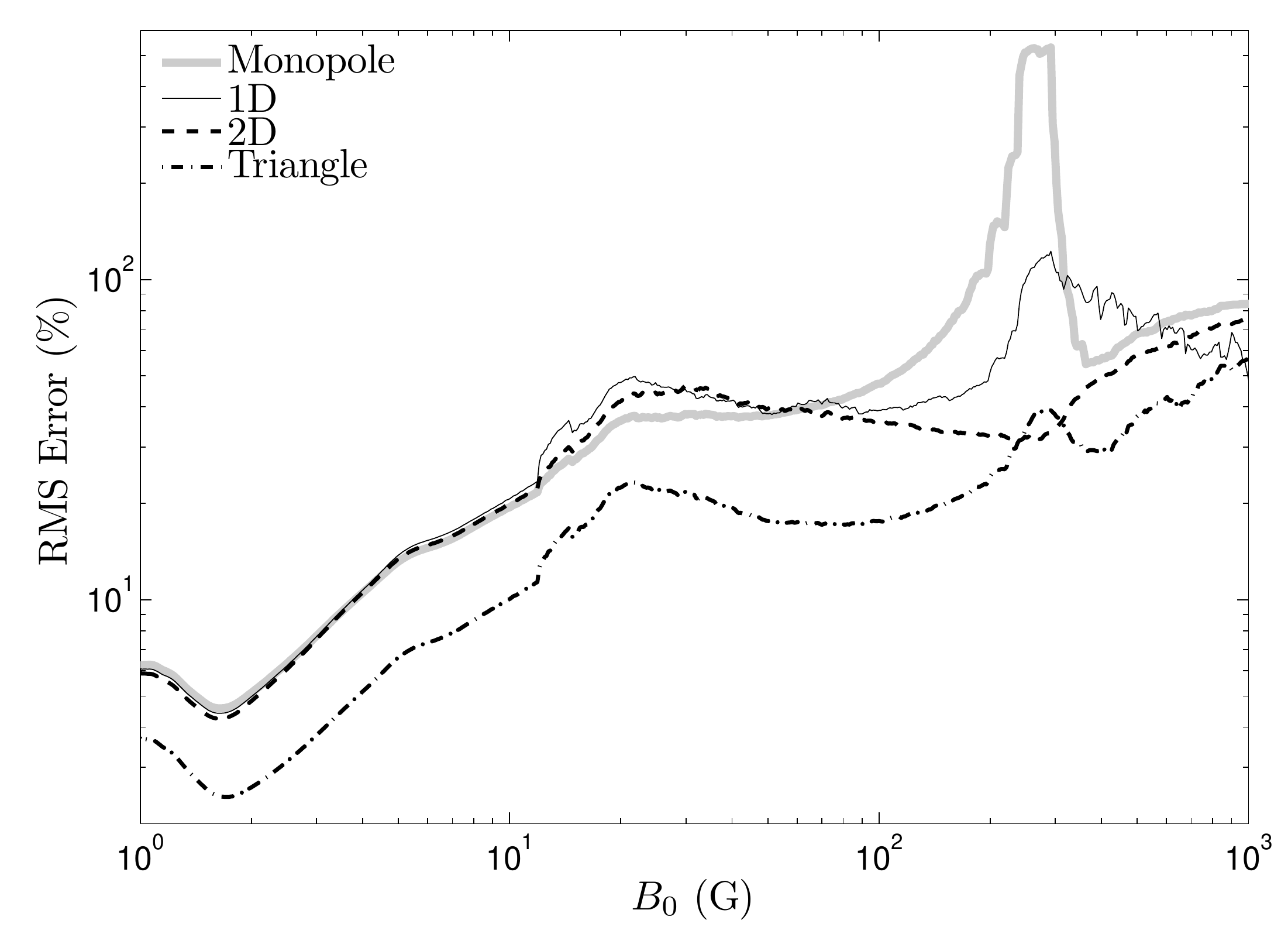}}
\caption{\label{fig:isocomparison}RMS accuracy of the field within regions centered around field isosurfaces above a transverse domain wall (200~nm $\times$ 5~nm wire). For a given $B_{\rm 0}$, the error is averaged over points with field within 10\% of $B_{\rm 0}$.}
\end{figure}

In an effort to present a figure of merit independent of sample size and distance from the wall center, we now perform analysis over regions of specific field strength. This removes biases due to changes in the volume of space examined, and will also be independent of nanowire size, whilst still assessing the fidelity of the models to the true field shape. The regions analyzed are shells centered on field isosurfaces. Whilst for some applications the field at a given distance is the focus, there are a number of applications, e.g. atom trapping \cite{nanotrap}, where the working regime is defined by the field strength itself. The following analysis illustrates the error across these different regimes of field. We compute the error over a region where, according to micromagnetic simulations, $0.9B_{\rm 0}\le|\vec{B}|\le 1.1B_{\rm 0}$, for some $B_{\rm 0}$. The results of this analysis are shown in Figure~\ref{fig:isocomparison}.
A familiar trend is observed; the error increases for larger fields (shorter distances), and the error decreases for models which more faithfully replicate the shape of real domain walls. For very large fields the differences between models become less clear. This is expected as the shape is not accurately reproduced by any of the models in the near field. Again there is convergence at low fields as the approximation of the domain wall as a point object becomes increasingly valid. The biggest benefit of using a more complex model is seen in the range $\sim$10-100~G. It is precisely this region where the shape of the field begins to reflect the asymmetry of the domain wall shape, but does not show the complex behaviour observed at very short distances.

Discussion up to this point has been based upon the field size. Similar analysis was also performed for the field direction. This has inherent difficulties as $B_x$ and $B_y$ have zero points at all $z$. By examining the angle between the computed field directions we observe a good level of accuracy. E.g. consider the error in the field direction at a given height above the domain wall, using the triangular model. For a 200~nm $\times$ 5~nm wire the mean (maximum) error is less than $6^{\circ}$ ($20^{\circ}$) for all heights \textgreater 80~nm. There is a significant reduction in accuracy at very short distance. This inaccuracy is, in some regions, stark. The reason for this is small regions of negative effective charge. For a transverse domain wall such a region is found at the point of the triangle. The resulting large error in $B_z$ is then responsible for the inaccuracy in field direction, which is prominent for $z<100$~nm; higher than this there is excellent agreement, as with the field magnitude.

The analysis presented here has been for one particular wire geometry. To examine the flexibility of our approach to model nanowires with other dimensions we also considered five other wire geometries, detailed in Table~\ref{tab:geometries}. Note that the three geometries with larger wire thicknesses correspond to wires which host vortex-type domain walls. For these wires the triangular model is no longer appropriate since, as can be seen from Figure~\ref{fig:dw}, the magnetic pole distribution for a vortex-type domain wall is rectangular in shape. A summary of the figures of merit for all models and wire geometries is provided in Appendix~\ref{sec:fom}.
\begin{center}
\begin{table}[h]\footnotesize
\caption{\label{tab:geometries}The six geometries of wire examined in comparisons with fields from micromagnetic methods.}
\begin{ruledtabular}\footnotesize
\begin{tabular}{cccc}
{\bf Label} & {\bf Width (nm)} & {\bf Thickness (nm)} & {\bf Wall type}\\
\hline
A & 100 & 5 & Transverse\\
B & 200 & 5 & Transverse\\
C & 400 & 5 & Transverse\\
D & 100 & 15 & Vortex\\
E & 200 & 15 & Vortex\\
F & 400 & 15 & Vortex\\
\end{tabular}
\end{ruledtabular}
\end{table}
\end{center}

A general trend is observed that there is a loss of accuracy with an increase in nanowire size, shown explicitly in Appendix~\ref{sec:fom}. This is due to larger wires having larger and more prominent `near field' regions. Whilst the figures presented only show one wire geometry, they are representative of all wire sizes; the distributions of error show a similar shape for all six wires considered.

The analysis presented has used the assumption $s=w$. In Appendix~\ref{sec:fom} we provide the RMS percentage error using an optimized value of $s$, $E_{\rm RMS}^{\prime}$. Comparing this with the RMS error using $s=w$, $E_{\rm RMS}$, it is clear that there is only a very small loss of accuracy when using this approximation.

\section{Conclusions}
A series of analytic models of fields from domain walls in nanomagnetic structures have been derived and developed to incorporate the characteristic shape of transverse-type walls. The results obtained from these models were compared with micromagnetic simulations. Improved accuracy was observed for models incorporating higher dimensionality in domain wall structure, reflecting the fact that domain walls are not simple point like objects. We have also shown that no \emph{a priori} knowledge is required that might limit the usefulness of these models; assuming the wall width (along the wire) to be equal to the wire width results in negligible loss of accuracy. 

Data from analytic models show better agreement at larger distances from the nanomagnetic structure -- regions where the field is smaller. This is intuitively expected as further from the domain wall the approximation of it being a point object is increasingly accurate.
At very short range more detailed structures are observed in micromagnetic simulations. The models presented in this paper do not reproduce this.

Examples have been presented where an RMS error over an extended region of interest of less than 5\% is achieved. For fields up to $\sim$100~G an error of less than 10\% can be achieved. The maximum field for a given height is reproduced accurately by all but the simplest of models. Similar accuracy is also obtained when considering the direction of the field.

The models presented give an efficient and intuitive way of calculating the fields from domain walls in nanowires. The loss of accuracy compared to detailed micromagnetic simulations is small. The methods provided are quicker and much more accessible than existing techniques.

\section*{Acknowledgements}
The authors gratefully acknowledge financial support from EPSRC under grants EP/F025459/1 and EP/F024886/1.

\appendix
\section{Fringing field expressions}
\label{sec:fieldexp}
Here we provide the derivations of the expressions for the magnetic field for all the presented models. The symbols used throughout are defined as follows: $M_{\rm s}$ = saturation magnetization, $\mu_{\rm 0}$ = permeability of free space, $w$ = wire/domain wall width, $t$ = wire/domain wall thickness, $s$ = domain wall length and
$\{x,y,z\}$ = spatial coordinates.

\subsection{Simple Monopole}
In the presence of a magnetic medium we have from Maxwell's equations:
\begin{equation}
\mu_0\vec{\nabla}.\left(\vec{H}+\vec{M}\right)=0,
\end{equation}
where $\vec{H}$ is the magnetic field and $\vec{M}$ the magnetization. Considering the wall as an extended object (Figure~\ref{fig:dwcharge}) with magnetic charge density $\rho_{\rm m}$ we have in analogy to Gauss' law for electrostatics
\begin{equation}
\vec{\nabla}.\vec{H}=\rho_{\rm m}/\mu_{\rm 0}.
\end{equation}
The `magnetic charge', $q_{\rm m}$, associated with this volume is then given by
\begin{equation}
q_{\rm m}=-\mu_{\rm 0}\int\vec{\nabla}.\vec{M}\ {\rm d}V=-\mu_{\rm 0}\int\vec{M}.\hat{n}\ {\rm d}S.\label{eq:surfint}
\end{equation}
Here we employed the divergence theorem; $\hat{n}$ is the unit normal of the surface element d$S$ with $S$ enclosing $V$. 

Assuming a discontinuous magnetization reversal the volume of charge tends to zero. This is illustrated in Figure~\ref{fig:dwcharge} by considering $s\to 0$. Evaluation of Equation~(\ref{eq:surfint}) yields the charge associated with a head-to-head wall,
\begin{equation}
\label{eq:charge}
q_{\rm m}=2\mu_{\rm 0}M_{\rm s}wt.
\end{equation}
A tail-to-tail wall has a charge of $-q_{\rm m}$. Equation~(\ref{eq:charge}) has a form similar to other expressions for effective magnetic charge, c.f.\ e.g. \cite{ladak}. The magnetic field at a position $\vec{r}$ is then given, in direct analogy to Coulomb's law, by
\begin{equation}
\vec{B}(\vec{r})=\frac{\mu_{\rm 0}M_{\rm s}wt}{2\pi\left|\vec{r}\right|^2}\hat{r}.
\end{equation}

\subsection{1D Model}
We now extend the domain wall from a point charge to a line of charge. Infinitesimal elements of this charge, d$q_{\rm m}$, contained in a length ${\rm d}x_{\rm N}$, are given by
\begin{equation}
\label{eq:dq1d}
{\rm d}q_{\rm m}=2\mu_{\rm 0}M_{\rm s}t\ {\rm d}x_{\rm N}.
\end{equation}
The contributions from across the entire wire width are summed in the following integral to give the components of magnetic field, $B_i$:
\begin{equation}
\label{eq:bi1d}
B_i=\frac{\mu_{\rm 0}M_{\rm s}t}{2\pi}\int_{-w/2}^{w/2}\frac{r_i}{\left|\vec{r}\right|^3}\ {\rm d}x_{\rm N},
\end{equation}
where we define $\vec{r}=(r_x,r_y,r_z)=(x-x_{\rm N},y-y_{\rm N},z)$ the vector from an infinitesimal charge element located at $(x_{\rm N},y_{\rm N},0)$ to the point under consideration ($y_{\rm N}=0$ in this case). This then yields the expressions given in Eqs.~(\ref{eq:1dbx})-(\ref{eq:1dbz})\footnote{An alternative route to deriving these expressions is to use the magnetic scalar potential and integrating in an entirely analogous manner; the magnetic field is then simply the negative of the gradient of the scalar potential. Whilst entirely equivalent, within this paper we integrate expressions for the magnetic field directly throughout for the sake of clarity.}.

\subsection{2D Rectangle}
The final extension to the model which we derive analytically is to consider the domain wall as a 2D object in the $xy$ plane, as in Figure \ref{fig:2db}. The wall size can now vary arbitrarily along the wire length, to have width $s$. We choose to fix the value of $s$ to be $w$, the wire width. We rescale ${\rm d}q_{\rm m}$ to reflect the change in domain wall size. By analogy to Equation~(\ref{eq:dq1d}) we have
\begin{equation}
\label{eq:dqrect}
{\rm d}q_{\rm m}=2\mu_{\rm 0}M_{\rm s}t/w\ {\rm d}x_{\rm N}{\rm d}y_{\rm N},
\end{equation}
which is the infinitesimal element of charge associated with an area element d$A=$d$x_{\rm N}$d$y_{\rm N}$. Using this expression the total field is then given by integrating over all d$A$.
\begin{equation}
\label{eq:dintrect}
B_i=\int_{-w/2}^{w/2}\int_{-w/2}^{w/2}\frac{\mu_{\rm 0}M_{\rm s}tr_i}{2\pi w\left|\vec{r}\right|^3}\ {\rm d}y_{\rm N}{\rm d}x_{\rm N}.
\end{equation}
Evaluation of these integrals was performed symbolically \cite{mathematica}, see also e.g. \cite{gradshteyn}.


\begin{widetext}
\section{Figures of merit}
\label{sec:fom}
The following table shows all the figures of merit for the six different wire sizes according to the various models.
\begin{center}
\begin{table}[h]\footnotesize
\caption{\label{tab:summary}A summary of the figures of merit. $E_{\rm RMS}$ is the RMS error over all points, $E_{\rm RMS}^{\prime}$ is the RMS error with optimal $s$, $E_{\rm M}$ is the mean percentage error, $E_{\rm RMS}^{\rm Max B}$ is the RMS error in the maximum field for a given height, $E_{\rm Max}$ is the maximum observed percentage error in the field; $E_{\rm Max}^{>100}$ is this value limited to heights \textgreater 100~nm, $\Delta B_{\rm Max}$ is the maximum observed error in the field; $\Delta B_{\rm Max}^{>100}$ is this value limited to heights \textgreater 100~nm. The labels A-F refer to the wire geometries detailed in Table~\ref{tab:geometries}.}
\begin{ruledtabular}\footnotesize
\begin{tabular}{@{}cp{15pt}@{}p{15pt}@{}p{15pt}@{}p{15pt}@{}p{15pt}@{}p{15pt}@{}p{15pt} @{}p{15pt}@{}p{15pt}@{}p{15pt}@{}p{15pt}@{}p{15pt}@{}p{15pt}@{}p{15pt}@{}p{15pt}@{}p{15pt} @{}p{15pt}@{}p{15pt}@{}p{15pt}@{}p{15pt}@{}p{15pt}@{}p{15pt}@{}p{15pt}@{}p{16pt}@{}p{18pt} @{}p{16pt}@{}p{16pt}@{}}
& \multicolumn{21}{c}{\underline{{\bf Model}}} \\
& \multicolumn{6}{c}{{\bf Monopole}} & \multicolumn{6}{c}{{\bf 1D}} & \multicolumn{6}{c}{{\bf 2D}} & \multicolumn{3}{c}{{\hspace{-10pt}\bf Triangle}} \\
\hline
{\bf Quantity} & \textbf{A} & \textbf{B} & \textbf{C} & \textbf{D} & \textbf{E} & \textbf{F} & \textbf{A} & \textbf{B} & \textbf{C} & \textbf{D} & \textbf{E} & \textbf{F} & \textbf{A} & \textbf{B} & \textbf{C} & \textbf{D} & \textbf{E} & \textbf{F} & \textbf{A} & \textbf{B} & \textbf{C}\\
\hline
$E_{\rm RMS}~(\%)$ & 11 & 20 & 67 & 6 & 21 & 129 & 11 & 18 & 35 & 5 & 17 & 59 & 10 & 16 & 30 & 4 & 10 & 26 & 7 & 8 & 14\\
$E_{\rm RMS}^{\prime}$ (\%) & - & - & - & - & - & - & - & - & - & - & - & - & 10 & 16 & 30 & 3 & 7 & 19 & 6 & 8 & 14\\
$E_{\rm M}~(\%)$ & 7 & 10 & 21 & 2 & 6 & 24 & 7 & 10 & 19 & 2 & 6 & 22 & 7 & 10 & 17 & 1 & 3 & 8 & 4 & 5 & 8\\
$E_{\rm RMS}^{\rm Max B}~(\%)$ & 110 & 169 & 355 & 116 & 330 & 720 & 15 & 10 & 14 & 27 & 52 & 85 & 3 & 4 & 10 & 7 & 10 & 15 & 3 & 6 & 12\\
$E_{\rm Max}$ ($\times 10^3$ \%) & 1.9 & 6.4 & 29 & 0.94 & 2.9 & 75 & 0.58 & 1.2 & 1.8 & 0.88 & 1.8 & 16 & 0.27 & 0.70 & 1.2 & 0.26 & 0.38 & 3.2 & 0.27 & 0.49 & 0.56\\
$E_{\rm Max}^{>100}$ (\%) & 60 & 86 & 330 & 47 & 198 & 890 & 66 & 103 & 191 & 31 & 116 & 377 & 38 & 76 & 124 & 21 & 63 & 164 & 38 & 43 & 61\\
$\Delta B_{\rm Max}$ ($\times 10^3$ G) & 8.2 & 17 & 34 & 23 & 50 & 100 & 1.3 & 1.4 & 1.6 & 4.4 & 4.8 & 5.0 & 0.38 & 0.84 & 0.97 & 1.3 & 1.3 & 1.6 & 0.36 & 0.65 & 0.88\\
$\Delta B_{\rm Max}^{>100}$ (G) & 25 & 78 & 264 & 82 & 343 & 928 & 21 & 40 & 77 & 55 & 192 & 358 & 16 & 38 & 63 & 32 & 99 & 142 & 11 & 16 & 43\\
\end{tabular}
\end{ruledtabular}
\end{table}
\end{center}
\vspace{-20pt}
\end{widetext}

\end{document}